\documentclass{article} 
\usepackage{amsmath}
\usepackage{amsfonts}
\usepackage{amssymb}
\usepackage{amscd}

\def\p{{\sf p}}

\def\x{{\sf x}}

\def\y{{\sf y}}

\begin{document}

\title{Progress in Euclidean relativistic few-body quantum mechanics
\footnote{Presented at LC2012, Krakow, Poland;
Research supported by the US DOE Office of Science}}
\author{Wayne Polyzou\\
The University of Iowa\\
Iowa City,  IA 52242}

\maketitle

\begin{abstract}
We discuss recent progress in the Euclidean formulation of 
relativistic few-body quantum mechanics.
\end{abstract}
  
\section{Introduction}

Euclidean relativistic quantum mechanics is a formalism for
constructing quantum-mechanical models that have a unitary
representation of the Poincar\'e group on a model Hilbert
space\cite{1}.  The advantage of this approach is that it is
straightforward to formulate relativistic few-body models that satisfy
cluster separability along with a spectral condition.  A distinctive
feature is that the dynamics is formulated in terms of truncated
Euclidean Green functions rather than a Hamiltonian with few-body
interactions.  We discus the construction of Green functions that
satisfy reflection positivity, which is a sufficient condition on the
truncated Green functions to ensure that the Hamiltonian satisfies a
spectral condition and for the positivity of quantum probabilities.
We also discuss conditions for establishing the existence of
scattering wave operators and discuss the computation of $S$-matrix
observables.

Below we briefly summarize the structure of the theory.  A dense set
of vectors in the model Hilbert space, ${\cal H}$, is represented by a
collection of functions of Euclidean space-time variables, $\x_i$,
that have support for positive relative Euclidean times:
\[
f \to (f_0, f_1 (\x) , f_2 (\x_1,\x_2) , \cdots ) 
\]
satisfying
\[
\mbox{\bf support of }\, f_k(\x_1,\cdots ,\x_k)  =  
\{
\x_1, \cdots ,\x_k \,\vert \, 0 <\x_1^0 < \x_2^0 < \x_3^0  <\cdots < \x_k^0   \}.
\]
The Euclidean time reflection operator, $\theta$, is defined by
$\theta (\x^0,\pmb{\x}) := (- \x^0,\pmb{\x})$.  The quantum mechanical
inner product is expressed in terms of a collection of Euclidean
invariant Green functions
\[
\{ G_{m:n} (\x_m,\cdots ,\x_1; \y_1, \cdots , \y_n) \}
\]
by 
\[
\langle f \vert g \rangle = ( f , \theta G g)_E := 
\]
\[
\sum_{m,n} \int f_m^* (\x_1, \cdots ,\x_m )
G_{m;n} (\theta \x_m,\cdots ,\theta \x_1; \y_1, \cdots , \y_n) g(\y_1, \cdots , \y_n) 
d^{4m} \x d^{4n}\y ,
\]
where the $\x_i$ variables are final variables and the $\y_i$
variables are initial variables.  The collection of Green functions
are called reflection positive when $\langle f \vert f \rangle \geq 0$
for all functions satisfying the positive relative-time support
condition.

The collection of Green functions satisfy cluster properties if  
\[
\lim_{\vert \mathbf{a}\vert \to \infty}
G_{m:n} (\x_m+\mathbf{a} ,\cdots , \x_{k+1}+\mathbf{a} ,\x_{k}, \cdots , \x_1;
\y_1+\mathbf{a} ,\cdots , \y_l+\mathbf{a} ,\y_{l+1}, \cdots , \y_n)  =
\]
\[
G_{k:l} (\x_k ,\cdots ,  \x_1 ; \y_1 ,\cdots ,  \y_l )
G_{m-k,n-l}(\x_{m} ,\cdots ,  \x_{k+1} ; \y_{l+1} ,\cdots ,  \y_{n} )
\]

Poincar\'e generators $\{ H, \mathbf{P}, \mathbf{J}, \mathbf{K} \}$ 
on ${\cal H}$ and defined by
\[
\langle \x \vert H \vert \mathbf{f} \rangle :=
\{0 ,{\partial \over \partial
\x^0_{11}} f_1 (\x_{11}), 
\left ( {\partial \over \partial
\x^0_{21}} + {\partial \over \partial
\x^0_{22}} \right )  f_2 (\x_{21},\x_{22}), \cdots  \}
\]

\[
\langle \x \vert \mathbf{P}  \vert \mathbf{f} \rangle :=
\{0 ,  - i {\partial \over
\partial \vec{\x_{11}}} f_1 (\x_{11}), 
 - i \left ( {\partial \over
\partial \vec{\x_{21}}}  + {\partial \over
\partial \vec{\x_{22}}} \right ) f_2 (\x_{21},\x_{22}), \cdots  \}
\]

\[
\langle \x \vert \mathbf{J}  \vert \mathbf{f} \rangle :=
\{0 , - i \vec{\x}_{11} \times
{\partial \over \partial \vec{\x}_{11}} f_1 (\x_{11}), 
\]
\[
  - i\left (  \vec{\x}_{21} \times
{\partial \over \partial \vec{\x}_{21}}  + \vec{\x}_{22} \times
{\partial \over \partial \vec{\x}_{22}} 
\right  ) f_2 (\x_{21},\x_{22}), \cdots  \}
\]

\[
\langle \x \vert \mathbf{K}  \vert \mathbf{f} \rangle :=
\{0 , \left ( \vec{\x}_{11} 
{\partial \over \partial \x_{11}^0}- \x_{11}^0  {\partial \over
\partial \vec{\x}_{11}} \right )  f_1 (\x_{11}),
\]

\[
\left ( \vec{\x}_{21} 
{\partial \over \partial \x_{21}^0}- \x_{21}^0  {\partial \over
\partial \vec{\x}_{21}}+ \vec{\x}_{22} 
{\partial \over \partial \x_{22}^0}- \x_{22}^0  {\partial \over
\partial \vec{\x}_{22}} \right )  f_2 (\x_{21},\x_{22}), \cdots  \}.
\]
These operators are Hermetian and satisfy the Poincar\'e commutation
relations on ${\cal H}$. 
 
The invariant mass and transfer matrix, which are dynamical operators,
are easily computed in this representation:
\[
\langle \x \vert e^
{-\beta H} \vert f \rangle =
\]
\[
(f_0, f_1 (\x^0-\beta,\pmb{\x}) , f_2 
(\x_1^0-\beta,\pmb{\x}_1,\x_2^0-\beta,\pmb{\x}_2) , \cdots ) \to
\]

\[
M^2 = ({\partial^2 \over \partial \beta^2} +
{\partial \over \partial \mathbf{a}}\cdot 
{\partial \over \partial \mathbf{a}} ) 
\langle \x \vert e^
{-\beta H - i\mathbf{a}\cdot \mathbf{P}} \vert f 
\rangle_{\vert_{\beta=0, \mathbf{a}=\mathbf{0}}}
 =
\]
\[
({\partial^2 \over \partial \beta^2} +
{\partial \over \partial \mathbf{a}}\cdot 
{\partial \over \partial \mathbf{a}} ) 
(f_0, f_1 (\x^0-\beta,\pmb{\x}-\mathbf{a}) , 
\]
\[
f_2 
(\x_1^0-\beta,\pmb{\x}_1-\mathbf{a},\x_2^0-\beta,\pmb{\x}_2-\mathbf{a}) , \cdots )_{\vert_{\beta=0, \mathbf{a}=\mathbf{0}}}.
\]
These properties are motivated by the Osterwalder-Schrader
reconstruction theorem of local field theory\cite{2}.  The difference
between the Green functions of a local field theory and few-body
quantum mechanics is that in the quantum mechanical case we only
retain a finite number of these functions. In addition, in local field
theory there is only one $N$-point Green function; while in the
quantum-mechanical case there may be different $N$-point Green
functions corresponding different designations of the initial and
final Euclidean space-time coordinates.  The full symmetry in the
local field theory case leads to crossing symmetry, which may be
violated in the quantum mechanical case.

The product, $\langle f \vert f \rangle$, is related to the standard
Minkowski-space inner product. This is illustrated in
the one-body case by the well-known \cite{3} calculation
\[
\langle f \vert f 
\rangle = 
\int f^*(x) G_{1:1}(\Theta x;y) f(y) d^4x d^4y 
\]
\[ 
= {1 \over (2 \pi)^4} \int d^4\x d^4\y d^4\p dm 
f^* (\x) { e^{i \p \cdot (\theta 
\x-\y)} \rho (m )
\over \p^2 + m^2}  f(\y)  
\]
\[
=\int {d^3 p dm \rho(m) \over 2\omega_{m} (\mathbf{p} \,) }
\vert g(\mathbf{p},m) \vert^2 
\geq 0
\]
where the Euclidean and Minkowski wave functions $f(\x)$ and
$g(\mathbf{p},m)$ are related by
\[
g(\mathbf{p},m) = {1 \over (2 \pi)^{3/2}} 
\int f(\x_0,\pmb{\x}) e^{-\omega_m(\mathbf{p})\x_0 -i \pmb{\x}\cdot 
\mathbf{p}} d^4\x
\]
and the Lorentz invariant measure. $d^3p/\omega_m(\mathbf{p})$,  
appears naturally. 

One of the challenges of constructing models based on the Euclidean
formulation of relativistic quantum mechanics is the problem of
finding a robust class of reflection-positive model Green functions.
Two-body truncated Green functions with standard K\"all\'en-Lehmann
representations are reflection positive, as illustrated above.  One
difficulty with multipoint Green functions is that reflection
positivity is not stable\cite{4} with respect to small Euclidean
invariant perturbations.  For example, if one starts with a product of
reflection-positive free Green function, and solves the Bethe-Salpeter
equation with a small Euclidean-invariant kernel for the four-point
Green function, the resulting Green function is not automatically
reflection positive\cite{4}.  On the other hand, Widder\cite{5}
demonstrated that the most general solution in 1 dimension to

\[
\int_0^\infty  f^*(t) g(t+t') f(t') dt dt'> 0 \
\]
has the form 
\[
g(t) = \int e^{-\lambda t} \rho(\lambda ) d\lambda  =
\int {\lambda \over \pi} {e^{itp} \over \lambda^2 + p^2} \rho(\lambda) dp d\lambda
\]
which has a structure similar to the K\"all\'en-Lehmann representation of the 
Euclidean Green function.
This observations suggest considering  
integral representations of connected four point Green functions of the form
\[
G^c_{2:2} (\x_2,\x_1;\y_1,\y_2) =
\]
\[
\int e^{i \p_1 \cdot (\x_2-\x_1)}
e^{i \p_2 \cdot (\x_1-\y_1)} e^{i \p_3 \cdot (\y_1-\y_2)} \times
\]
\begin{equation}
{g(\p_1,\p_2,\p_3 ,m_2 )\over
(\p_1^2 + m^2) (\p_2^2 + m_2^2 )
(\p_3^2 + m^2)} d^4\p_1 d^4\p_2 d^4\p_3 dm_2 . 
\label{a}
\end{equation}
Calculations show that this class of Green functions are reflection
positive subject to mild conditions on $g(\p_1,\p_2,\p_3 ,m_2 )$.  It
is straightforward to generalize this to higher order connected Green
functions.  What simplifies the reflection-positivity constraint,
compared to the field theory case, is that the truncated four point
Green functions $G_{1:3}(\x_1:\x_2,\x_3,\x_4)$,
$G_{2:2}(\x_1,\x_2:\x_3,\x_4)$, and 
$G_{3:1}(\x_1,\x_2,\x_3;\x_4)$ do not have to be related.  In the
local field theory case they must be identified, which leads to
additional restrictions on $g(\p_1,\p_2,\p_3 ,m_2 )$.

Another complication is the formulation of scattering theory. 
This is because the dynamics enters in the structure of the
Hilbert space inner product, so there is no asymptotic dynamics.  In
addition, the real-time evolution operator is difficult to construct
in this formalism, while the transfer matrix involves a simple
quadrature.  The absence of an asymptotic dynamics can be treated
using the two-Hilbert space formulation of scattering\cite{6}.
This requires solving the one-body problem for subsystems.  In this
form, time-dependent methods can be used to define scattering wave
operators and a simple generalization of Cook's method can be used to
test the existence of the wave operators.

The first step is to solve the mass eigenvalue problem  
\[ 
\langle \x \vert (M^2-\lambda^2) \vert \lambda \rangle  = 0 
\]
for eigenfunctions in the pure point spectrum of $M^2$ associated with
a subsystem Green function.  Here $\langle \x \vert$ is a shorthand
notation for $\langle \x_1 \cdots \x_m\vert$ which are the initial or
final variables of the subsystem Green function.

Next translations and rotations are used to extract sharp momentum
and spin eigenstates of the same mass
\[
\langle \x \vert \lambda , \mathbf{p} \rangle 
 = 
\int {d^3a \over (2 \pi)^{3/2}}  e^{-i \mathbf{p}\cdot \mathbf{a}}  
\langle \x-\mathbf{a}  \vert \lambda \rangle 
\]
\[
\langle \x \vert \lambda,j, \mathbf{p},\mu  \rangle =  
\int_{SU(2)} dR  \sum_{\nu=-j}^j 
\langle \x \vert \lambda , R^{-1}\mathbf{p} \rangle 
D^{j*}_{\mu \nu} (R).
\]
These one-particle solutions are used to construct a map 
from an asymptotic Hilbert space to 
the physical Hilbert space by taking symmetrized products of the 
``one-particle'' plane-wave eigenstates 
\[
\langle \x \vert \Phi \vert \mathbf{p}_1, \mu_1, \cdots \mathbf{p}_k ,
\mu_k \rangle =  
\prod_i  \langle \x_{i_1} \cdots \x_{i_{n_i}} \vert 
\lambda_i , j_i, \mathbf{p}_i, \mu_i \rangle .
\]
Wave operators are defined by 
\[
\vert \Psi_{\pm} (g_1, \cdots g_n) \rangle : = 
\lim_{t \to \infty} e^{iHt} \Phi e^{-i H_0t} \vert \mathbf{g} \rangle =
\Omega_{\pm} \vert \mathbf{g} \rangle
\]
where $\vert \mathbf{g} \rangle $ represents wave packets in the
asymptotic particles' momenta and spin and $H_0 = \sum_i 
\omega_{m_i}(\mathbf{p}_i)$. 

A sufficient condition for the existence of this limit is the  
Cook condition\cite{7} 
\begin{equation}
\int_0^{\pm \infty} \Vert (H\Phi-\Phi H_0) e^{-i H_0t} \vert \mathbf{g} 
\rangle \Vert dt 
< \infty .
\label{b}
\end{equation}
For $N=2$ with $G_4 = G_2 G_2 + G^c_4$ the $G_2 G_2$ contribution to
$\Vert (H\Phi-\Phi H_0) e^{-i H_0t} \vert \mathbf{g} \rangle \Vert $
vanishes.  What remains is a regularity condition that depends only on
the truncated four-point function, $G^c_4$.  It is interesting note
that the truncated reflection-positive four point 
Euclidean Green functions in \ref{a} are distributions rather than 
short-ranged kernels; however
when one computes the integrand in (\ref{b}), it becomes a localized
kernel after integrating over the relative $p^0$ energy variables.

To calculate $S$-matrix elements the invariance principle \cite{8}
can be used, which allows us to make the replacement 
\[
H \to w(H) \qquad w(H) = -e^{-\beta H} \qquad \beta > 0 
\]
in the limits used to define the $S$ matrix elements
\[
S= \lim_{n \to \infty}  
\langle \mathbf{g}_f \vert e^{-in e^{-\beta H_0} } \Phi^{\dagger}
e^{2in e^{-\beta H} } 
\Phi e^{-i n e^{-\beta H_0} } \vert \mathbf{g}_i \rangle 
\]
Because the spectrum of $e^{-\beta H}$ is compact, for any fixed $n$,
$e^{2in e^{-\beta H} }$ can be uniformly approximated by a polynomial
in $e^{-\beta H}$. Recall that these matrix elements are related to
the transfer matrix, $\langle f \vert T(0,n\beta) \vert g\rangle =
\langle f \vert e^{-n \beta H} \vert g \rangle$, which can be
calculated using only quadratures.  Test calculations
\cite{1} demonstrate that this method can be used to
accurately calculate GeV-scale scattering cross sections.


\end{document}